# Image Improvement in Linear-Array Photoacoustic Imaging using High Resolution Coherence Factor Weighting Technique

**Moein Mozaffarzadeh**[a,b,c]**, Mohammad Mehrmohammadi**[d]**, Bahador Makkiabadi**[a,b,*]

[a]Department of Medical Physics and Biomedical Engineering, School of Medicine, Tehran University of Medical Sciences, Tehran, Iran.
[b]Research Center for Biomedical Technologies and Robotics (RCBTR), Institute for Advanced Medical Technologies (IAMT), Tehran, Iran.
[c]Department of Biomedical Engineering, Tarbiat Modares University, Tehran, Iran
[d]Department of Biomedical Engineering, Wayne State University, 818 W Hancock Street, Detroit, MI 48201.

**Abstract.** In Photoacoustic imaging (PAI), the most prevalent beamforming algorithm is delay-and-sum (DAS) due to its simple implementation. However, it results in a low quality image affected by the high level of sidelobes. Coherence factor (CF) can be used to address the sidelobes in the reconstructed images by DAS, but the resolution improvement is not good enough compared to the high resolution beamformers such as minimum variance (MV). As a weighting algorithm in linear-array PAI, it was proposed to use high-resolution-CF (HRCF) weighting technique in which MV is used instead of the existing DAS in the formula of the conventional CF. The higher performance of HRCF was proved numerically and experimentally. The quantitative results obtained with the simulations show that at the depth of 40 $mm$, in comparison with DAS+CF and MV+CF, HRCF improves the full-width-half-maximum of about 91 % and 15 % and the signal-to-noise ratio about 40 % and 14 %, respectively. Moreover, the contrast ratio at the depth of 20 $mm$ has been improved about 62 % and 21 % by HRCF, compared to DAS+CF and MV+CF, respectively.

*Bahador Makkiabadi, b-makkiabadi@tums.ac.ir

## 1 Introduction

In photoacoustic imaging (PAI), a short electromagnetic pulse, i.e. laser or radio frequency (RF), illuminates the target of imaging, and Ultrasound (US) waves are generated based on the thermoelastic effect.[1,2] In comparison with other imaging modalities, PAI has multiple advantages leading to many investigations.[3,4] The main incentive in PAI is having the merits of the US imaging spatial resolution and the optical imaging contrast in one imaging modality.[5] PAI can be used in different fields of study such as tumor detection,[6,7] ocular imaging,[8] monitoring oxygenation in blood vessels[9] and functional imaging.[10,11] Moreover, contrast agents and nanoparticles play a significant role in PAI.[12,13] PAI can be separated into two fields: photoacoustic tomography (PAT) and photoacoustic microscopy (PAM).[14,15] PAT, for the first time, was successfully used as *in vivo* functional and structural brain imaging modality in small animals.[16] In PAT, an array of elements may be formed in linear, arc or circular shape, and mathematical reconstruction algorithms are used to obtain the optical absorption distribution map of the tissue.[17–19] Most of the used reconstruction algorithms for image formation in PAI are based on some assumptions leading to artifacts and disturbing effects on the formed photoacoustic (PA) images. One of the challenges in PA image formation is related to reduction of these effects for different number of transducers and properties of imaging media.[20,21]

Image formation in US imaging and PAI is similar, and most of the algorithms in US image formation can be used in PAI.[22,23] However, there would be some modifications if an algorithm in US imaging is going to be used in PAI. These modifications have led using different hardware to implement an integrated US-PA imaging device. Many studies focused on developing one beamforming technique for US and PA image formation in order to reduce the cost of the imaging systems.[24,25] DAS is the most commonly used beamforming algorithm in PAI. However, it leads to a low quality image, having a wide mainlobe and high level of sidelobes.[26] Adaptive beamformers, commonly employed in radar, have the ability of weighting the aperture based on the characteristics of detected signals, providing a high quality image with a wide range of off-axis signals rejection. MV can be treated as one of the commonly used adaptive methods in medical US imaging.[27–29] Vast variety of modifications have been investigated on MV such as complexity reduction,[30,31] shadowing suppression,[32] using eigenstructure to enhance MV performance,[33,34] and combination of MV and multi-line transmission (MLT) technique.[35] Matrone *et al.* proposed a new algorithm namely delay-multiply-and-sum (DMAS), as a beamforming technique for medical US imaging.[36] This algorithm, introduced by Lim *et al.* , was initially used in the confocal microwave imaging for breast cancer detection.[37] DMAS was used in synthetic aperture imaging.[38]

Double stage DMAS (DS-DMAS), outperforming DMAS in the terms of contrast and sidelobes, was introduced for the linear-array PAI.[39,40] Minimum variance-based DMAS (MVB-DMAS) has been proposed for resolution improvement in DMAS while the level of sidelobes would be retained.[41,42] Its eigenspace-based version is also investigated.[43,44] Coherence factor (CF) can be mentioned as one of the prevalent weighting methods in beamforming field.[45] The performance of CF has been investigated for US imaging and PAI in[28] and,[40] respectively. Moreover, high resolution CF (HRCF) has been investigated for high-frame rate US imaging.[46,47] Recently, a modified version of the CF has been reported by the authors.[48]

In this paper, the performance of HRCF is investigated for linear-array PAI. The concept of this technique indicates that a high resolution image obtained with an algorithm such as MV can be used for weighting the calculated samples instead of the formed image by DAS. It is shown that the proposed weighting algorithm, used with DAS, outperforms DAS and MV (with/without CF) in the terms of resolution, sidelobes and contrast.

The rest of the paper is organized as follows. In section 2, the concept of beamforming and the proposed method are presented. Numerical and experimental results are presented in section 3 and 4, respectively. The advantages and disadvantages of the proposed method are discussed in section 5. Finally, the conclusion is presented in section 6.

## 2 Materials and Methods

In this section, the concept of image reconstruction in linear-array PAI, along with the concerned algorithms in this paper, are discussed.

*2.0.1 Beamforming*

In linear-array PAI, a laser illuminates the target of imaging. Then, PA signals are recorded using an US transducer. The detected signals can be used for the image formation using a beamforming algorithm. The most common beamforming algorithm in linear-array PAI is DAS. Its formula is as follows:

$$y_{DAS}(k) = \sum_{i=1}^{M} x_i(k - \Delta_i), \quad (1)$$

where $y_{DAS}(k)$ is the output of the beamformer, $k$ is the time index, $M$ is the number of elements of array, and $x_i(k)$ and $\Delta_i$ are the detected signals and the corresponding time delay for the detector $i$, respectively. To have a more efficient beamformer and improve the reconstructed image, CF can be combined with DAS.[28] The combination of DAS and CF results in sidelobes reduction and contrast enhancement. CF, as an effective weighting process, is given by:

$$CF(k) = \frac{\left| \sum_{i=1}^{M} x_i(k - \Delta_i) \right|^2}{M \sum_{i=1}^{M} \left| x_i(k - \Delta_i) \right|^2}. \quad (2)$$

As can be seen in (2), the numerator is the output of DAS algorithm. (1) can be simply implemented and provides a real-time PAI. However, due to the low range of the off-axis signals rejection, it leads to low quality images. The combination of DAS and CF can be written as follows:

$$y_{DAS+CF}(k) = CF(k) * y_{DAS}(k). \quad (3)$$

MV can be chosen as an algorithm which provides a high resolution in PAI.[41] However, sidelobes caused by MV highly affect the image quality and degrade the contrast of the reconstructed image.

The output of MV adaptive beamformer is given by:

$$y(k) = \boldsymbol{W}^H(k)\boldsymbol{X}_d(k) = \sum_{i=1}^{M} w_i(k)x_i(k - \Delta_i), \quad (4)$$

where $\boldsymbol{X}_d(k)$ is the time-delayed array detected signals $\boldsymbol{X}_d(k) = [x_1(k), x_2(k), ..., x_M(k)]^T$, $\boldsymbol{W}(k) = [w_1(k), w_2(k), ..., w_M(k)]^T$ is the beamformer weights, and $(.)^T$ and $(.)^H$ represent the transpose and the conjugate transpose, respectively. The detected array signals can be written as follows:

$$\boldsymbol{X}(k) = \boldsymbol{s}(k) + \boldsymbol{i}(k) + \boldsymbol{n}(k) = s(k)\boldsymbol{a} + \boldsymbol{i}(k) + \boldsymbol{n}(k), \quad (5)$$

where $\boldsymbol{s}(k), \boldsymbol{i}(k)$ and $\boldsymbol{n}(k)$ are the desired signal, interference and noise components received by the transducer, respectively. Parameters $s(k)$ and $\boldsymbol{a}$ are the signal waveform and the related steering vector, respectively. MV bemaformer can be used to adaptively weight the calculated samples. Its goal is to achieve the optimal weights in order to estimate the desired signal as accurately as possible. The superiority of MV algorithm has been evaluated in comparison with static windows, such as Hamming window.[29] To acquire the optimal weights, signal-to-interference-plus-noise ratio ($SINR$) needs to be maximized:[49]

$$SINR = \frac{\sigma_s^2|\boldsymbol{W}^H\boldsymbol{a}|^2}{\boldsymbol{W}^H\boldsymbol{R}_{i+n}\boldsymbol{W}}, \quad (6)$$

where $\boldsymbol{R}_{i+n}$ and $\sigma_s^2$ are the $M \times M$ interference-plus-noise covariance matrix and the signal power, respectively. The maximization of $SINR$ can be gained by minimizing the output interference-plus-noise power while maintaining a distortionless response to the desired signal using following

equation:

$$\min_{\boldsymbol{W}} \boldsymbol{W}^H \boldsymbol{R}_{i+n} \boldsymbol{W}, \quad s.t. \quad \boldsymbol{W}^H \boldsymbol{a} = 1. \tag{7}$$

The solution of (7) is given by:[50]

$$\boldsymbol{W}_{opt} = \frac{\boldsymbol{R}_{i+n}^{-1} \boldsymbol{a}}{\boldsymbol{a}^H \boldsymbol{R}_{i+n}^{-1} \boldsymbol{a}}. \tag{8}$$

In practice, the interference-plus-noise covariance matrix is unavailable. Consequently, the sample covariance matrix is used instead of the unavailable covariance matrix using $N$ recently received samples and is given by:

$$\hat{\boldsymbol{R}} = \frac{1}{N} \sum_{n=1}^{N} \boldsymbol{X}_d(n) \boldsymbol{X}_d(n)^H. \tag{9}$$

Using MV in medical US imaging encounters some problems which are addressed and discussed in reference.[28] The subarray-averaging or the spatial-smoothing method can be used to achieve a better estimation of the covariance matrix using decorrelation of the coherent signals received by the array. The covariance matrix estimation using the spatial-smoothing can be written as:

$$\hat{\boldsymbol{R}}(k) = \frac{1}{M-L+1} \sum_{l=1}^{M-L+1} \boldsymbol{X}_d^l(k) \boldsymbol{X}_d^l(k)^H, \tag{10}$$

where $L$ is the subarray length, and $\boldsymbol{X}_d^l(k) = [x_d^l(k), x_d^{l+1}(k), ..., x_d^{l+L-1}(k)]$ is the delayed input signal for the $l_{th}$ subarray. Due to the limited statistical information, only a few temporal samples are used to estimate the covariance matrix. Therefore, to obtain a stable covariance matrix, the diagonal loading ($DL$) technique is used. This method leads to replacing $\hat{\boldsymbol{R}}$ by the loaded sample covariance matrix, $\hat{\boldsymbol{R}}_l = \hat{\boldsymbol{R}} + \gamma \boldsymbol{I}$, where $\gamma$ is the loading factor:

$$\gamma = \Delta.trace\{\hat{\boldsymbol{R}}(k)\}, \tag{11}$$

where $\Delta$ is a constant related to the subarray length. Also, the temporal averaging method can be applied along with the spatial averaging to gain resolution enhancement while the contrast is retained. The estimation of the covariance matrix using both temporal averaging and spatial smoothing in given by:

$$\hat{\boldsymbol{R}}(k) = \frac{1}{(2K+1)(M-L+1)} \times \sum_{n=-K}^{K} \sum_{l=1}^{M-L+1} \boldsymbol{X}_d^l(k+n)\boldsymbol{X}_d^l(k+n)^H, \qquad (12)$$

where the temporal averaging is performed over $(2K+1)$ samples. After estimation of the covariance matrix, the optimal weights are calculated by (8) and (12). Finally, the output of MV beamformer is given by:

$$\hat{y}_{MV}(k) = \frac{1}{M-L+1} \sum_{l=1}^{M-L+1} \boldsymbol{W}_*^H(k)\boldsymbol{X}_d^l(k). \qquad (13)$$

where $\boldsymbol{W}_*(k) = [w_1(k), w_2(k), ..., w_L(k)]^T$. Considering (2), it can be seen that the numerator of the fraction is the output of DAS beamformer, and this is why the output of the combination of DAS and CF does not have a high resolution. To put it more simply, the combination of DAS and CF does not provide a high resolution because DAS is weighted using a procedure in which DAS plays a significant role. On the other hand, using MV combined with CF weighting is a good alternative. However, as will be shown in the next section, the output of the combination of DAS and MV can be further improved using HRCF weighting procedure combined with DAS. Its formula is as follows:[46,47]

$$y_{DAS+HRCF}(k) = HRCF(k) * y_{DAS}(k), \qquad (14)$$

where

$$HRCF(k) = \frac{M\left|\hat{y}_{MV}(k)\right|^2}{\sum_{i=1}^{M}\left|x_i(k-\Delta_i)\right|^2}. \quad (15)$$

In the next section, the results of the proposed method (the combination of DAS and HRCF) for PA image reconstruction is evaluated.

## 3 Numerical Results and Performance Assessment

In this section, numerical results are presented to illustrate the performance of the proposed technique for PA image formation in comparison with DAS, DAS+CF, MV and MV+CF.

### 3.1 Simulated Point Targets

#### 3.1.1 Simulation Setup

The K-wave Matlab toolbox was used to simulate the numerical study.[51] Imaging region was 20 $mm$ in the lateral axis and 80 $mm$ in the vertical axis. A linear-array having $M$=128 elements operating at 7 $MHz$ central frequency and 77% fractional bandwidth was used to detect the PA signals generated from the defined initial pressures. The speed of sound was assumed to be 1540 $m/s$ during the simulations. The sampling frequency was 50 $MHz$, subarray length $L$=$M$/2, $K$=5 and $\Delta = 1/100L$ for all the simulations.

#### 3.1.2 Qualitative and Quantitative Evaluation

The reconstructed images are shown in Fig 1, along with a zoomed version at the depth of 40 $mm$ (shown in Fig 2) for a better evaluation. As can be seen, the reconstructed image using DAS have a low quality along with high sidelobes. MV improves the resolution significantly, but the sidelobes

Table 1: FWHM ($\mu m$) in -6 $dB$ values at the different depths.

| Depth($mm$) \ Beamformer | DAS | DAS+CF | MV | MV+CF | DAS+HRCF |
|---|---|---|---|---|---|
| 25 | 1106 | 677 | 118 | 118 | 102 |
| 30 | 1388 | 848 | 127 | 126 | 104 |
| 35 | 1632 | 986 | 130 | 130 | 105 |
| 40 | 1942 | 1179 | 121 | 121 | 103 |
| 45 | 2284 | 1388 | 131 | 131 | 106 |
| 50 | 2684 | 1619 | 138 | 137 | 108 |
| 55 | 3068 | 1862 | 144 | 144 | 110 |

still affect the image. Using CF combined with DAS or MV results in sidelobes reduction and image quality enhancement. Even though the image reconstructed by MV+CF, shown in Fig 1(d), has a high resolution, but the negative effects of the sidelobes still degrade the image quality. In Fig 1(e), it can be seen that the sidelobes are reduced compared to Fig 1(d) while the resolution is retained. To assess in more details, the lateral variations of the reconstructed images shown in Fig 1 are shown at four imaging depths in Fig 3. As can be seen, DAS+HRCF results in lower level of sidelobes and narrower width of mainlobe compared to other beamformers. Moreover, the lateral valleys between the targets have the lowest levels using the proposed method. Consider, for instance, the depth of 50 $mm$ where the level of sidelobes are of about -36 $dB$, -69 $dB$, -45 $dB$, -79 $dB$ and -88 $dB$ for DAS, DAS+CF, MV, MV+CF and DAS+HRCF, respectively. Thus, the proposed method leads to lowest sidelobes in comparison with other beamformers. Moreover, the levels of the lateral valleys for DAS, DAS+CF, MV, MV+CF and DAS+HRCF are about -29 $dB$, -61 $dB$, -37 $dB$, -70 $dB$ and -80 $dB$, respectively. It indicates the higher separability of the proposed method. To evaluate the proposed method quantitatively, the full-width-half-maximum (FWHM) in -6 $dB$ and signal-to-noise ratio (SNR) metrics are calculated and presented in Table 1 and Table 2, respectively. SNRs are calculated using the formula explained in.[39] As can be seen in Table 1, the proposed method for PA image reconstruction results in a narrower width of

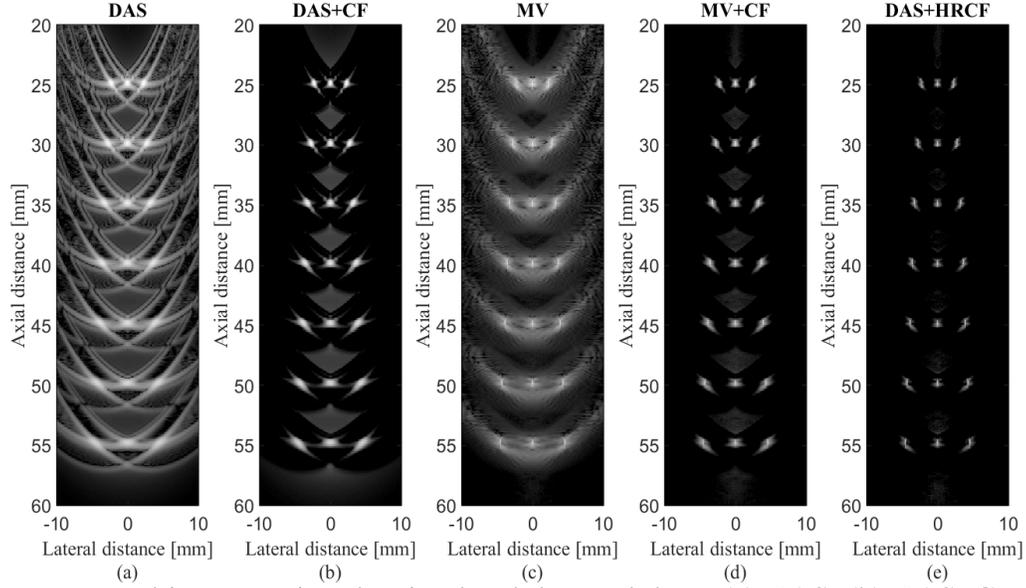

Fig 1: Reconstructed images using the simulated detected data. (a) DAS, (b) DAS+CF, (c) MV, (d) MV+CF and (e) DAS+HRCF. A linear-array and point phantom were used for the numerical design. All images are shown with a dynamic range of 60 $dB$. Noise was added to the detected signals having a SNR of 40 $dB$.

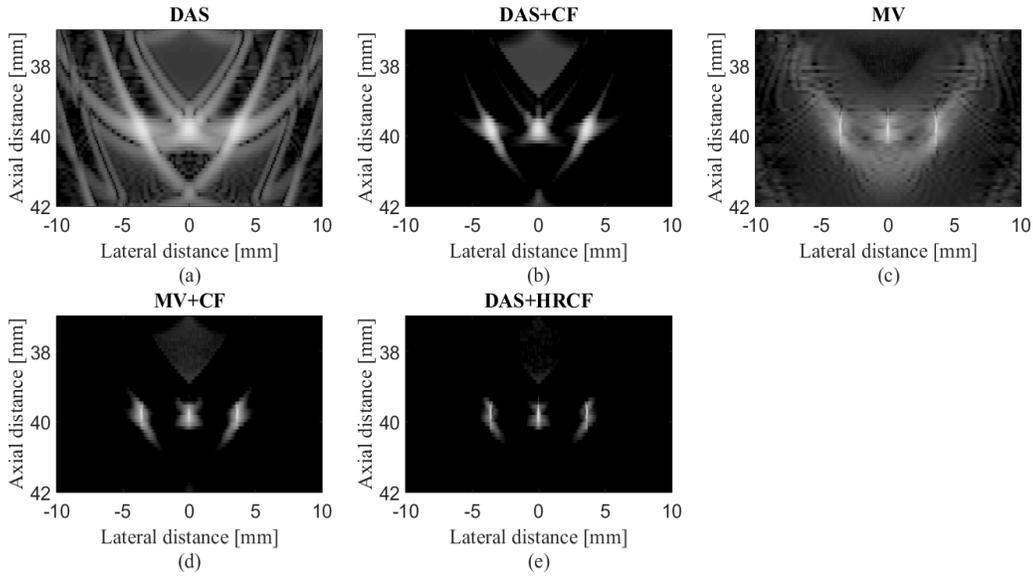

Fig 2: A close view of the reconstructed images shown in Fig 1.

mainlobe in -6 $dB$ compared to other beamformers in the all depth of imaging. Of note, there is no significant improvement compared to MV and MV+CF. Consider, in particular, the depth of 45 $mm$ where DAS, DAS+CF, MV, MV+CF and DAS+HRCF leads to a FWHM of 2284 $\mu m$, 1388

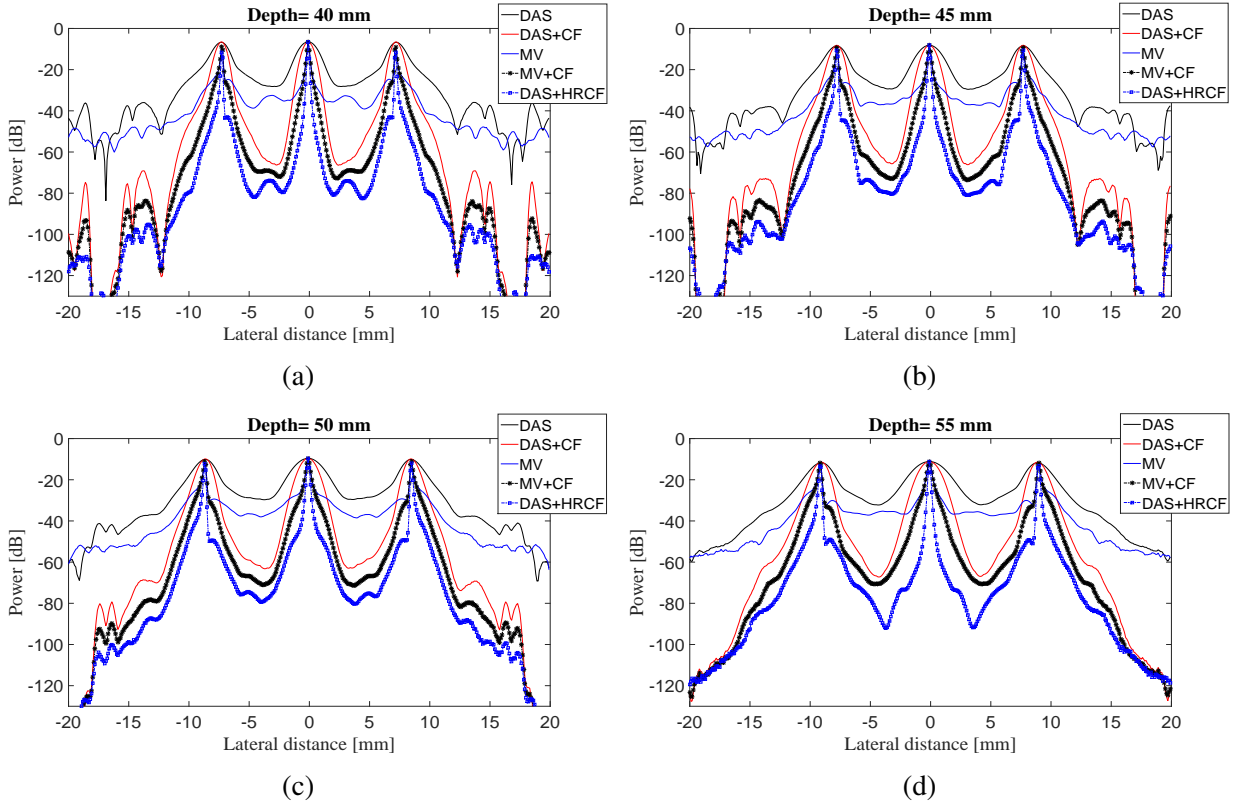

Fig 3: Lateral variations of the reconstructed images shown in Fig 1 at the depths of (a) 20 $mm$, (b) 40 $mm$, (c) 55 $mm$ and (d) 70 $mm$.

Table 2: SNR ($dB$) values at the different depths.

| Depth($mm$) \ Beamformer | DAS | DAS+CF | MV | MV+CF | DAS+HRCF |
|---|---|---|---|---|---|
| 25 | 48.9 | 66.8 | 59.8 | 80.6 | 90.9 |
| 30 | 46.6 | 64.5 | 57.2 | 77.9 | 87.9 |
| 35 | 44.2 | 62.0 | 54.8 | 75.4 | 85.4 |
| 40 | 42.4 | 60.2 | 53.2 | 73.9 | 84.0 |
| 45 | 40.8 | 58.5 | 51.2 | 71.8 | 81.6 |
| 50 | 39.3 | 57.2 | 49.5 | 70.32 | 79.8 |
| 55 | 37.8 | 55.4 | 47.8 | 68.3 | 77.7 |

$\mu m$, 131 $\mu m$, 131 $\mu m$ and 103 $\mu m$, respectively. In comparison with a high resolution method such as MV, the proposed method leads to 28 $\mu m$ FWHM improvement. As shown in Table 2, the proposed method results in a higher SNR in comparison with other reconstruction methods at the all depths of imaging. Consider, for instance, the depth of 55 $mm$ where DAS, DAS+CF, MV, MV+CF and DAS+HRCF results in a SNR of 37.8 $dB$, 55.4 $dB$, 47.8 $dB$, 68.3 $dB$ and 77.7 $dB$, respectively. In other words, DAS+HRCF improves the SNR for about 9 $dB$ and 22 $dB$ compared to MV+CF and DAS+CF, respectively, proving its superiority for linear-array PAI.

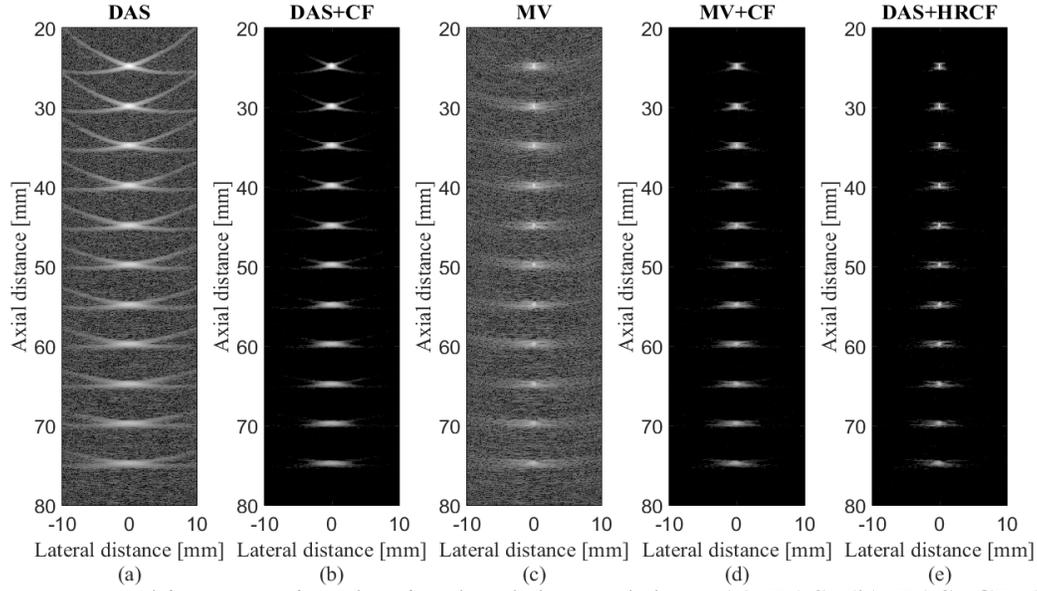

Fig 4: Reconstructed images using the simulated detected data. (a) DAS, (b) DAS+CF, (c) MV, (d) MV+CF and (e) DAS+HRCF. A linear-array and point phantom were used for the numerical design. All images are shown with a dynamic range of 60 $dB$. Noise was added to the detected signals having a SNR of 0 $dB$.

The proposed method is evaluated at the presence of high level of imaging noise. Eleven 0.1 $mm$ radius spherical absorbers as initial pressure were positioned along the vertical axis every 5 $mm$ beginning 25 $mm$ from transducer surface. Noise was added to the detected signals having a SNR of 0 $dB$. The reconstructed images are shown in Fig 4. As can be seen, the presence of the noise in the reconstructed images using DAS and MV degrade the images. CF results in the higher noise suppression and higher image quality, as shown in Fig 4(b) and Fig 4(d). As shown in Fig 4(e), the sidelobes are better reduced using DAS+HRCF.

The lateral variations for images shown in Fig 4, at two depths of imaging, are shown in Fig 5. As can be seen, the proposed method results in lower level of sidelobes and narrower width of mainlobe.

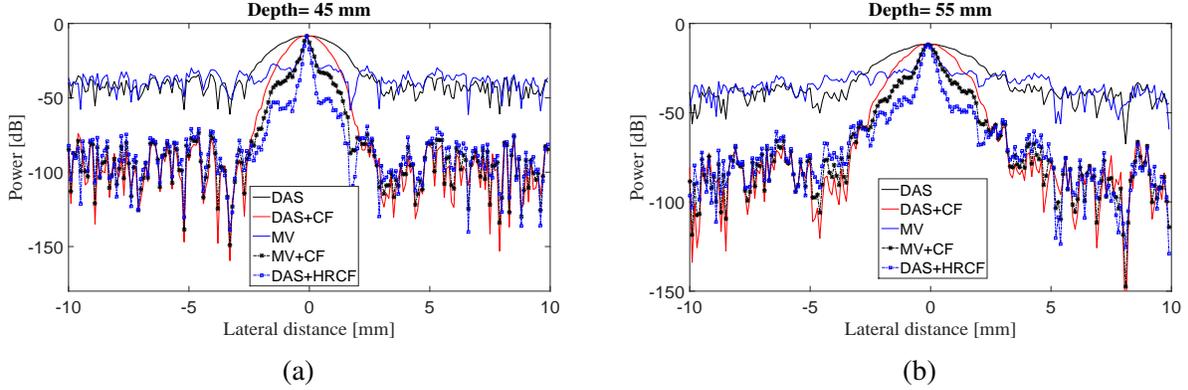

Fig 5: Lateral variations of the reconstructed images shown in Fig 4 at the depths of (a) 45 $mm$ and (b) 55 $mm$.

*3.1.3 Sensitivity to Sound Velocity*

In this section, the proposed method is evaluated in the term of robustness against the sound velocity errors. It results from the medium heterogeneity which are inevitable in practical imaging. The simulation design for Fig 4 is used in order to investigate the robustness, except that the sound velocity is overestimated by 5%, which covers and may be more than the typical estimation error in practice.[27,28] As can be seen in Fig 6, the effect of the overestimated sound velocity degrade the reconstructed image using DAS and MV. However, using CF the images are improved in the terms of presence of the noise and artifacts. Fig 6(d) shows that CF combined with MV leads to a high resolution and low sidelobes, but the sidelobes presence in the images still degrade the image quality. Comparing Fig 6(d) and Fig 6(e), it can be seen that using the proposed method the lower sidelobes in comparison with MV+CF can be achieved. The lateral variations shown in Fig 7 prove the superiority of DAS+HRCF compared to other beamformers.

*3.2 Simulated Cyst Targets*

Imaging region was 20 $mm$ in the lateral axis and 30 $mm$ in the vertical axis. Two massive cysts having a radius of 4 $mm$ have been positioned at the depths of 20 $mm$ and 29 $mm$.

*3.2.1 Qualitative and Quantitative Evaluation*

The reconstructed images using the beamformers are shown in Fig 8. As can be seen, DAS leads to a low quality image. Although MV improves the resolution, it causes high sidelobes and artifact. The CF enhances the image quality with reduction of sidelobes and artifact. As shown in Fig 8(e), DAS+HRCF results in well-separated cyst targets, having a high resolution and low sidelobes. To compare the images in details, contrast ratio (CR) has been calculated using the formula presented in,[39] and the Table 3 shows the measurements. As can be seen, CF significantly improves the CR, DAS+HRCF provides higher CR compared to other beamformers for the both massive cyst targets.

The proposed method has been evaluated at the presence of high level of imaging noise, having a SNR of 0 $dB$. The reconstructed images and the calculated CRs are shown in Fig 9 and Table 4, respectively. As can be seen in Fig 9, the added noise completely affects the reconstructed images

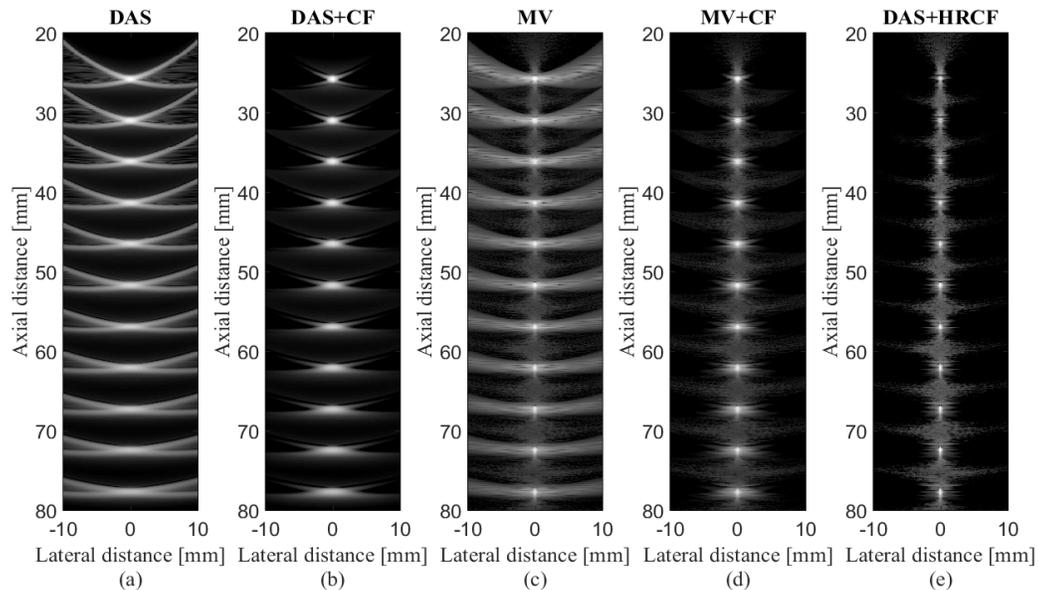

Fig 6: Reconstructed images using the simulated detected data. (a) DAS, (b) DAS+CF, (c) MV, (d) MV+CF and (e) DAS+HRCF. A linear-array and point phantom were used for the numerical design. All images are shown with a dynamic range of 60 $dB$. Noise was added to the detected signals having a SNR of 40 $dB$. The sound velocity was overestimated for 5% in the image reconstruction procedure.

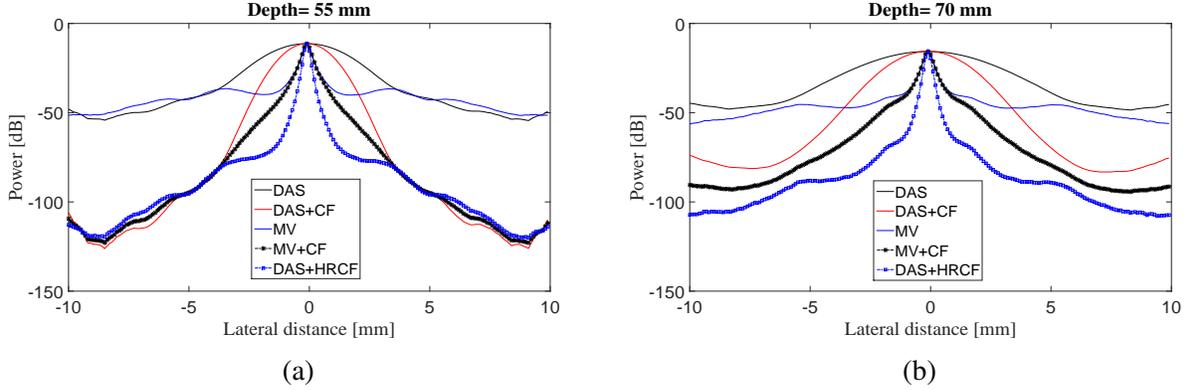

Fig 7: Lateral variations of the reconstructed images shown in Fig 4 at the depths of (a) 45 $mm$ and (b) 55 $mm$.

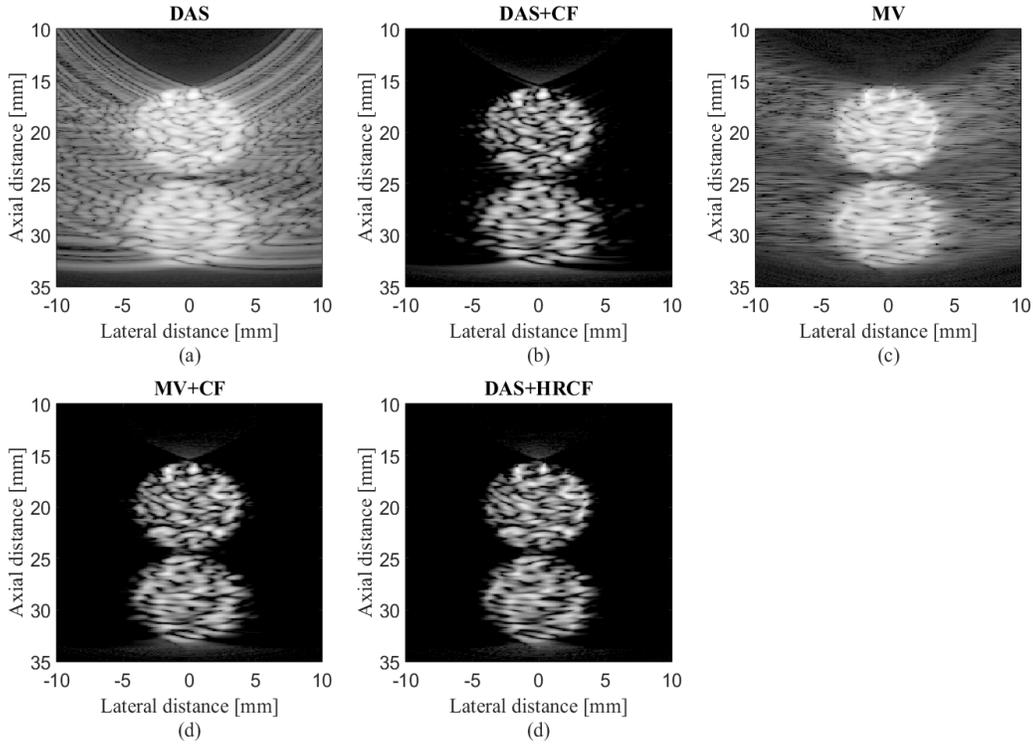

Fig 8: Reconstructed images using the simulated detected data. (a) DAS, (b) DAS+CF, (c) MV, (d) MV+CF and (e) DAS+HRCF. A linear-array and cyst phantom were used for the numerical design. All images are shown with a dynamic range of 60 $dB$. Noise was added to the detected signals having a SNR of 40 $dB$.

by DAS and MV, while using CF reduces the effects of the noise. The higher resolution of the proposed method, compared to MV+CF, is visible regarding the boundaries of the cyst located at the depth of 20 $mm$ ( see Fig 9(d) and Fig 9(e)). The calculated CR, shown in Table 4, indicates

Table 3: CR ($dB$) values at the different depths for images shown in Fig 8.

| Depth($mm$) \ Beamformer | DAS | DAS+CF | MV | MV+CF | DAS+HRCF, |
|---|---|---|---|---|---|
| 20 | 11.5 | 30.3 | 20.9 | 40.3 | 49.1 |
| 29 | 11.1 | 28.3 | 19.2 | 37.4 | 45.8 |

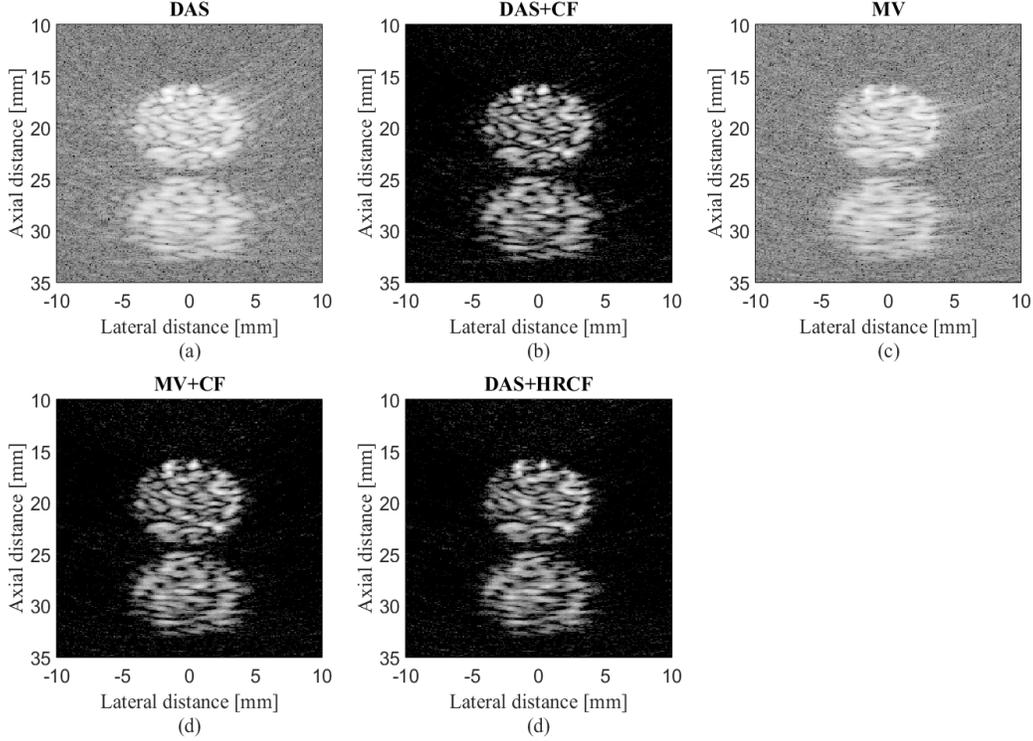

Fig 9: Reconstructed images using the simulated detected data. (a) DAS, (b) DAS+CF, (c) MV, (d) MV+CF and (e) DAS+HRCF. A linear-array and cyst phantom were used for the numerical design. All images are shown with a dynamic range of 60 $dB$. Noise was added to the detected signals having a SNR of 0 $dB$.

that DAS+HRCF outperforms other beamformers and provides a higher contrast.

## 4 Experimental Results

### 4.1 Experimental Setup

To further evaluate the proposed weighting method and its effects on enhancing the PA images, phantom experiments were performed in which a phantom consists of 2 light absorbing wires with diameter of 150 $\mu m$ were placed 12 $mm$ apart from each other 10. In this experiment, we utilized a Nd:YAG pulsed laser, with the pulse repletion rate of 30 $Hz$ at wavelengths of 532 $nm$. A programmable digital ultrasound scanner (Verasonics Vantage 128), equipped with a linear array transducer (L11-4v) operating at frequency range between 4 to 9 $MHz$ was utilized to acquire the PA RF data. A high speed FPGA was used to synchronize the light excitation and PA signal acquisition.

Table 4: CR ($dB$) values at the different depths for images shown in Fig 9.

| Depth($mm$) \ Beamformer | DAS | DAS+CF | MV | MV+CF | DAS+HRCF, |
|---|---|---|---|---|---|
| 20 | 10.0 | 26.7 | 13.6 | 32.2 | 36.6 |
| 29 | 9.7 | 26.3 | 13.2 | 30.5 | 33.6 |

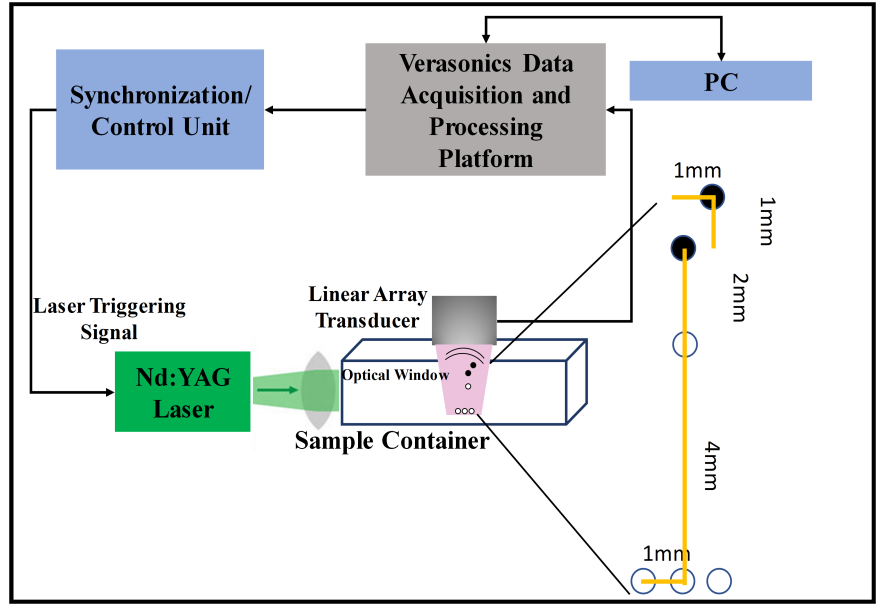

Fig 10: The schematic of the setup used for the experimental PAI.

*4.2 Qualitative and Quantitative Evaluation*

The reconstructed images are shown in Fig 11. As can be seen, the artifact and noise affect the reconstructed image by DAS while the CF improves the image quality by suppressing them. As shown in Fig 11(c), MV results in an image having a high resolution, but the presence of the noise highly affects the image. As can be seen in Fig 11(e), HRCF results in a high resolution while the sidelobes are degraded, and the presence of the noise is clearly lower than other methods, comparing the background of the Fig 11(e) with other images shown in Fig 11. To assess the images in details, the lateral variations of the two wire targets are shown in Fig 12. As can be seen, the HRCF outperforms the conventional CF combined with DAS and MV and results in a narrower width of mainlobe and lower level of sidelobes. Consider, for instance, the depth of 24 $mm$ where DAS+CF, MV+CF and DAS+HRCF result in -36 $dB$, -47 $dB$ and -60 $dB$ sidelobes, respectively. In other words, DAS+HRCF improves the sidelobes for about 24 $dB$ and 13 $dB$ compared to DAS+CF and MV+CF, respectively.

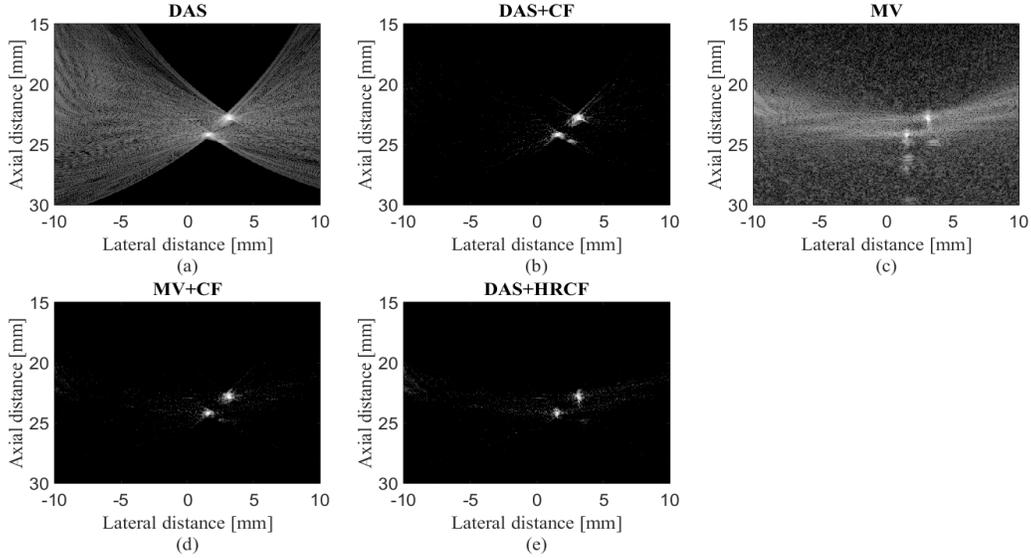

Fig 11: Reconstructed images using the experimental detected data. (a) DAS, (b) DAS+CF, (c) MV, (d) MV+CF and (e) DAS+HRCF. A linear-array and wire target phantom were used for the experimental design. All images are shown with a dynamic range of 60 $dB$.

Table 5: SNR ($dB$) values at the different depths for images shown in Fig 11.

| Beamformer<br>Depth($mm$) | DAS | DAS+CF | MV | MV+CF | DAS+HRCF, |
|---|---|---|---|---|---|
| 22 | 47.9 | 57.6 | 41.0 | 54.1 | 60.2 |
| 24 | 46.6 | 56.2 | 40.1 | 53.0 | 59.1 |

To compare the experimental images quantitatively, SNRs for all the methods are calculated and presented in Table 5 where the proposed weighting method leads to a higher SNR, for both imaging targets, compared to other methods, indicating the superiority of HRCF weighting method.

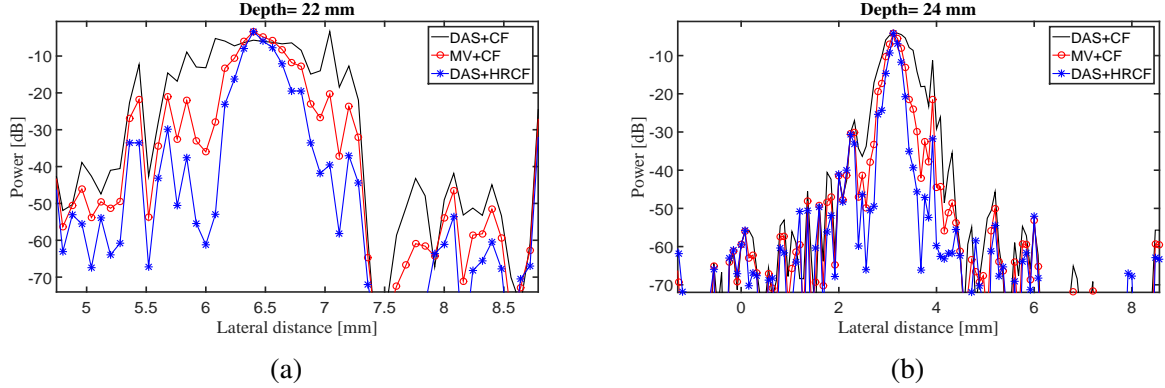

Fig 12: Lateral variations of the reconstructed images obtained with DAS+CF, MV+CF and DAS+HRCF, shown in Fig 11, at the depths of (a) 22 $mm$ and (b) 24 $mm$.

## 5  Discussion

The main improvement gained by HRCF is having a high resolution and low sidelobes at the same time. DAS is the most commonly used beamformer in PA and US imaging which is mainly as a result of its simple implementation. Moreover, it provides a real-time imaging. However, it results in a low quality image having a low resolution and high sidelobes due to its blindness and non-adaptiveness. To put it more simply, DAS considers all the calculated samples the same as each other, and there is just a summation process. On the other hand, adaptive beamformers, such as MV, provides a higher image quality compared to DAS, especially in the term of resolution. However, in MV, sidelobes affect the reconstructed image and degrade the image quality. CF is a weighting method that can be used with beamformers, such as DAS or MV, for sidelobes reduction. However, conventional CF weighting does not improve the resolution and the width of mainlobe significantly, compared to beamformers such as MV. It can be seen that in (2), the numerator of the formula of CF is the output of DAS. While CF reduces the sidelobes, the performance of CF is not high in the term of resolution, which is mainly due to the existence of DAS on the numerator of the formula of CF. Using MV instead of the exiting DAS in the (2) can improve the resolution gained by the conventional CF (15). The reconstructed images show that the HRCF outperforms CF combined with DAS and MV. As shown in Fig 1 and Fig 2, the point targets are better distinguished and detectable using HRCF weighting procedure, and the sidelobes are better reduced. The proposed method was evaluated in the terms of the presence of high level of noise and robustness to overestimated sound velocity, and the reconstructed images are shown in Figure 4 and Fig 6. As can be seen, the HRCF reduces the negative effects of the added noise, and it provides a higher robustness compared to other methods. The images have been evaluated using the lateral variations shown in Figure 3, Figure 5 and Figure 7, indicating the superiority of HRCF in the terms of sidelobes, lateral valley and the width of mainlobe. HRCF has been evaluated under the cyst targets, having two levels of noise, and it can be seen that the boundaries are sharpened, having a better detectable cysts as a result of the higher contrast of HRCF. Table 1, Table 2, Table 3 and Table 4 show the quantitative evaluation of the proposed weighting method. They indicate that the HRCF reduces the presence of the noise and results in the narrower width of mainlobe. Despite all the evaluation with the simulations, the algorithm should be evaluated using experimental data. The generated experimental images are shown in Figure 11, and the superiority of HRCF can be clearly seen. The lateral variations of the experimental images are shown in Figure 12, proving the higher performance of HRCF. It should be mentioned that the higher performance of the HRCF is obtained at the expense of the higher computational burden where replacing DAS by MV on the numerator of the formula of CF would result the order of complexity change from $O(M)$ to $O(L^3)$. All the results indicate that the HRCF can be an effective weighting method for image formation in linear-array PAI, and it provides a higher contrast and resolution compared to DAS and MV combined with conventional CF.

## 6 Conclusion

In this paper, the HRCF was proposed as a weighting method in linear-array PAI. It was shown that there is a DAS on the numerator of the formula of CF, and it can be replaced with MV beamformer. The proposed method (HRCF) was evaluated numerically and experimentally, and it was shown that it leads to a higher image quality compared to MV and DAS (with/without CF). The quantitative results show that at the depth of 55 $mm$, compared to DAS+CF and MV+CF, HRCF improves the SNR of about 9 $dB$ and 22 $dB$, respectively, and reduces the FWHM of about 1752 $\mu m$ and 44 $\mu m$, respectively.

**First Author** is an assistant professor at the University of Optical Engineering. He received his BS and MS degrees in physics from the University of Optics in 1985 and 1987, respectively, and his PhD degree in optics from the Institute of Technology in 1991. He is the author of more than 50 journal papers and has written three book chapters. His current research interests include optical interconnects, holography, and optoelectronic systems. He is a member of SPIE.

Biographies and photographs of the other authors are not available.

# List of Figures

## List of Tables